# Blockchain-Integrated Privacy-Preserving Medical Insurance Claim Processing Using Homomorphic Encryption

Diya Mamoria[1]    Harshit Jain[1]   Aswani Kumar Cherukuri[1*]

[1]School of Computer Science Engineering & Information Systems, Vellore Institute of Technology, Vellore, India.
[*]Email:  cherukuri@acm.org

*Abstract*—   This research proposes a decentralized and cryptographically secure framework to address the most acute issues of privacy, data security, and protection in the ecosystem of medical insurance claim processing. The scope of this study focuses on enabling the management of insurance claims in a transparent, privacy-protecting manner while maintaining the efficiency and trust level needed by the patients, healthcare providers, and insurers. To accomplish this, the proposed system adds blockchain technology to provide an unchangeable, decentralized, and auditable claim transactions ledger which enhances overall claim-related processes and trust among all stakeholders. To protect critical patient information, the framework employs homomorphic encryption—a modern form of cryptography—to allow authorized insurance providers to perform necessary operations like claim adjudication and reimbursement on encrypted medical records without any decryption during the process. This method significantly reduces the third-party processing privacy risk because patient data can be kept secret even when third-party processing is done. In addition, smart contracts improve automation of the most important procedures in the claim processing pipeline, which decreases manual, operational, and susceptibility towards human blunders or deceitful acts. The integration of these two transformative technologies—blockchain and homomorphic encryption—represents the core contribution of this work, enabling the coexistence of transparency and privacy which are usually viewed as competing objectives in traditional systems. The study additionally offers a prototype realization that illustrates the working possibility of the design together with some performance assessment results that prove its security, efficiency, and robustness under realistic conditions. As a result, these technologies are expected to foster the creation of a reliable, effective, and privacy safeguarding architecture that could transform the medical claim submission systems paradigm.

## I. INTRODUCTION

As the modern digital economy continues to evolve, the healthcare sector emerges with one of the highest levels of data generation and utilization as it handles countless personal and medical details on a day to day basis. This information contains a variety of highly sensitive components such as medical history, treatment descriptions, results of diagnostic tests, insurance claim records, and other financial dealings pertaining to patient care. These data sets form the core of the healthcare functioning and are ideal for the accuracy and precision of medical services. The breadth of this information raises several concerns, as maintaining confidentiality and integrity is essential for protecting patients' rights and ensuring institutional accountability.  The use of Electronic Health

Records (EHR), digital insurance claim processing systems, and cloud-based data have all contributed to the healthcare industry's recent fast digitization. These modifications have improved data-driven medical decision-making while also making healthcare faster and more accessible. [1]. However, these innovations have increased the risks posed to sensitive health information including cyber threats, system weaknesses, and inappropriate exploitation. To protect sensitive data from these new challenges, laws like the United States Health Insurance Portability and Accountability Act (HIPAA), GDPR were designed to safeguard health information from unauthorized disclosure, alteration, and access during its entire lifecycle.[2].

Healthcare data breaches continue to happen at a startling and frequently financially catastrophic regularity in spite of these rules. In addition to jeopardizing personal health information, these breaches undermine public confidence in healthcare organizations, highlighting the urgent need for stronger, privacy-preserving technical solutions that go beyond compliance to guarantee end-to-end security. The medical insurance claim processing pipeline is one of the healthcare ecosystem's most vulnerable and dangerous elements. Traditionally, this procedure relies on third-party intermediaries and centralized databases to manage, validate, and resolve claims. Although this centralized approach has long been the norm in the industry, it also brings a number of inefficiencies and security flaws. One of the most prominent examples highlighting the risks of this architecture is the Anthem data breach in 2015, during which hackers compromised the personal records of over 78 million individuals, including both patients and employees, exposing the fragility of centralized healthcare systems [3]. In addition to the inherent risks of centralized storage, research conducted by McLeod and Dolezel further reveals that many healthcare organizations face ongoing difficulties in implementing comprehensive and adaptive cybersecurity measures. These shortcomings create critical security gaps across the healthcare infrastructure, making systems more vulnerable to exploitation by malicious actors, whether external hackers or internal threats [4]. These combined factors make clear the urgent need for decentralized, transparent, and privacy-preserving solutions that can reinforce trust and security in the medical insurance claim processing environment.

*1.1 Background of Data Privacy Issues in Healthcare Insurance Systems*

Patients, healthcare providers, insurance companies, and third-party administrators are just a few of the stakeholders that must constantly and dynamically communicate personal health information (PHI) in order for the medical insurance industry to function. Validating insurance claims, processing reimbursements, and promoting efficient communication throughout the healthcare service chain all depend on this data interchange. However, sharing, storing, and transmitting PHI invariably creates a number of points of vulnerability where clever external attackers, malevolent insiders, or even unintentional system misconfigurations can take advantage of flaws and jeopardize patient confidentiality and data integrity [5]. Moreover, the rise of data-driven approaches, particularly the adoption of advanced data analytics and artificial intelligence (AI) algorithms in insurance claim adjudication, has amplified privacy concerns. These technologies often require access to large volumes of historical and real-time health data to detect patterns, predict fraud, and automate claim decisions. When such datasets are reused, shared across departments, or combined with external information sources, there is a heightened risk that even anonymized data can be reverse-engineered or re-identified, inadvertently exposing sensitive patient details [6]. While traditional de-identification techniques — including k-anonymity, l-diversity, and differential privacy — have been developed to mitigate these risks and obscure identifiable patterns, these methods are not foolproof. In particular, their effectiveness diminishes when datasets are subjected to cross-referencing against auxiliary external information, which can lead to successful re-identification attacks despite prior anonymization [7] . This growing gap between existing privacy protection mechanisms and the complexities of real-world healthcare data environments highlights the need for more advanced solutions, especially cryptographic techniques and decentralized system architectures, to minimize reliance on trust, mitigate vulnerabilities, and strengthen overall data security.

*1.2 Challenges in Claim Processing: Lack of Trust, Data Tampering, and Privacy Concerns*

The process of medical insurance claim handling has long been criticized for its inherent opacity, complexity, and inefficiency. Within the current system, the lack of clear and real-time transparency between key stakeholders — namely patients, healthcare providers, and insurance companies — frequently leads to the breakdown of trust. This disconnect can manifest in various operational issues, such as processing delays, administrative bottlenecks, and disputes over claim authenticity and coverage terms. Such inefficiencies often translate into a frustrating experience for patients and a costly, resource-draining cycle for healthcare institutions and insurers alike [8].

Compounding this issue is the structural reliance on centralized systems, which are typically responsible for maintaining and validating sensitive medical and financial data related to insurance claims. This centralization, while convenient for data consolidation and administrative control, introduces significant vulnerabilities. It create single points of failure that can be exploited through both external cyberattacks and internal misconduct. Whether through deliberate data tampering, unauthorized access, or accidental inconsistencies, the centralized architecture poses an ever-present risk to the integrity and reliability of medical insurance claim records.

In addition to systemic weaknesses, data fraud has emerged as a particularly pressing and costly concern within the medical insurance industry. Fraudulent activities — including false claim submissions, deliberate manipulation of billing codes, and the forging of medical documents — have become increasingly sophisticated and difficult to detect. These practices inflict substantial financial losses on the healthcare sector, amounting to billions of dollars in avoidable expenses every year [9].

At the same time, patients face an equally troubling dilemma. They are frequently required to disclose highly sensitive medical information during the claim verification process, often without receiving adequate clarity regarding how their data will be stored, accessed, or shared by insurance companies and associated third parties [10]. This lack of transparency not only erodes patient confidence but also raises significant ethical questions and places many traditional systems in direct conflict with contemporary data protection principles. International regulations, most notably the European Union's General Data Protection Regulation (GDPR), now emphasize the critical importance of safeguarding personal data and enforcing user consent in all data handling practices [11]. However, the fragmented and outdated architecture of many insurance claim systems continues to fall short of these evolving global privacy standards.

*1.3 The Role of Blockchain and Homomorphic Encryption in Solving These Challenges*

Emerging technologies such as blockchain and homomorphic encryption (HE) present a promising paradigm shift in the way secure data sharing, verification, and processing are approached within the healthcare insurance ecosystem. Blockchain, by design, functions as a decentralized, tamper-resistant ledger that is distributed across a network of participants, offering the potential to establish immutable audit trails that strengthen transparency, improve accountability, and significantly reduce reliance on a single point of control or failure [12]. In the specific context of medical insurance claims, blockchain technology can be leveraged to securely log claim transactions, maintain synchronized and shared access to claim histories for all authorized stakeholders, and enforce automated validation through smart contracts, which can transparently govern the rules of claim approval and payment settlements [13].

Smart contracts are particularly valuable in this process, as they allow claims to be automatically assessed and validated based on predefined eligibility conditions, thereby minimizing the need for manual oversight, reducing administrative delays, and limiting opportunities for fraudulent manipulation [14]. At the same time, homomorphic encryption complements this approach by enabling mathematical operations to be performed directly on encrypted data. This allows insurers to conduct essential claim computations — such as eligibility verification and reimbursement calculations — without ever accessing or decrypting sensitive patient information, thus preserving end-to-end data confidentiality throughout the process [15]. The combined application of these two technologies offers not only robust privacy and data integrity safeguards but also enhances overall operational efficiency and cultivates improved trust among all participating entities in the medical insurance claim environment.

*1.4 Problem Statement*

The existing medical insurance claim processing systems expose patients to significant privacy risks due to the need for full data disclosure during claim adjudication. Centralized system architectures are vulnerable to both internal and external data breaches and manipulation. Additionally, the lack of transparency and trust among patients, healthcare providers, and insurance companies leads to inefficiencies, delays, and elevated administrative costs. There is a clear need for a secure, decentralized, and privacy-preserving solution to address these systemic shortcomings.

*1.5 Objectives of the work*

In order to guarantee privacy, security, and operational effectiveness, this project attempts to create a blockchain-integrated medical insurance claim processing system strengthened by homomorphic encryption. The project's design includes a number of important goals. In order to lessen dependency on centralized authority and minimize single points of failure, it first aims to create a blockchain-based decentralized architecture that would allow transparent, tamper-resistant storage and management of medical insurance claim data. Second, the project intends to put into practice a homomorphic encryption technique that enables insurance companies to verify claims and carry out required calculations directly on encrypted medical records, protecting sensitive patient data all along the way. Third, it plans to employ smart contracts to automate various portions of the claim processing workflow, eliminating human participation, decreasing administrative overhead, and enhancing the speed and consistency of claim validation. Through a number of experimental validations and real-world case studies, the research will also test the system's durability and applicability as well as its performance, security, and privacy resilience. Lastly, the project hopes to contribute significant insights to the expanding field of secure healthcare informatics systems by showcasing the technological viability and benefits of incorporating homomorphic encryption and blockchain technology into medical insurance workflows.

## II. LITERATURE REVIEW

*2.1 Medical Insurance Claim Processing Systems*

Medical insurance claim processing forms a core and indispensable component of modern healthcare systems, where the accuracy, security, and efficiency in handling sensitive patient data are crucial to ensuring fair and timely service. The process typically involves multiple entities, including hospitals, clinics, insurance providers, third-party administrators, and in some cases, regulatory bodies — all interacting within a largely centralized infrastructure designed to manage medical and financial records.

In the traditional workflow, healthcare providers submit detailed medical records and invoices to insurance companies, which are then reviewed and verified against the terms of the patient's policy before any reimbursement is approved. While this process appears standardized in theory, it often proves to be cumbersome and prone to error in practice, owing to manual validation, human oversight, and inconsistencies in record formats. As the complexity of medical treatments grows and the volume of insurance claims steadily increases, these inefficiencies not only cause delays in legitimate reimbursements but also create frustration for both patients and healthcare providers, ultimately undermining trust in the entire system [16].

Beyond inefficiency, concerns over the security and privacy of sensitive patient information are especially pronounced in conventional insurance claim systems. Centralized storage solutions — widely adopted by insurers and healthcare providers — have long been criticized for their susceptibility to data breaches, insider threats, and unauthorized access. Since claims require the exchange of personal health information (PHI), patients often lose control over their data once it is submitted, raising ethical questions and compliance challenges. Even with standards like HIPAA, the persistence of data breaches and fragmented, non-interoperable systems continues to expose gaps in the protection of medical claim records [17]. These issues signal the growing need for more secure, transparent, and privacy-preserving solutions within the insurance claim processing landscape.

*2.2 Blockchain in Healthcare*

Blockchain technology has emerged in recent studies as a promising solution for addressing longstanding challenges in healthcare data management and insurance claim workflows. With its decentralized and tamper-resistant architecture, blockchain eliminates reliance on centralized authorities, allowing stakeholders such as patients, healthcare providers, insurers, and auditors to maintain synchronized and secure copies of shared data [18]. This structure significantly enhances data integrity, as each transaction — including treatment updates, diagnosis records, and claim

submissions — is cryptographically signed, time-stamped, and permanently recorded, minimizing risks of fraud and unauthorized alterations.

Smart contracts further strengthen blockchain's role in this domain by automating claim validation, payment processing, and dispute resolution based on predefined rules. This automation reduces administrative overhead and human error while improving efficiency and consistency in claim management [19]. In addition, blockchain's transparency allows patients to retain control over access permissions, reinforcing both data sovereignty and trust.

Existing literature highlights blockchain's successful applications across healthcare, including electronic health records, pharmaceutical supply chains, consent management, and secure data sharing, all of which demonstrate its capacity to re-engineer security and transparency in medical insurance claim processing [20].

*2.3 Homomorphic Encryption (HE)*

While blockchain addresses the integrity and traceability of healthcare data, privacy concerns persist, especially in scenarios where data needs to be computed or analyzed by third parties, such as insurance providers. This is where homomorphic encryption (HE) offers significant advantages over traditional encryption techniques. HE allows computations to be performed directly on encrypted data without the need for decryption, ensuring that sensitive patient information remains concealed from all processing entities throughout the claim verification pipeline [21]. This property is particularly important in medical insurance systems, where insurers must analyze patient records and treatment histories to determine claim eligibility and compute reimbursements, all while respecting stringent privacy regulations and ethical boundaries.

HE schemes can be broadly categorized into Partial Homomorphic Encryption (PHE), which supports unlimited operations of a single mathematical type (either addition or multiplication), Somewhat Homomorphic Encryption (SHE), which allows a limited number of both additions and multiplications, and Fully Homomorphic Encryption (FHE), which allows unlimited, arbitrary computations on encrypted data [22]. The concept of FHE, introduced by Gentry [15], has since evolved from theoretical frameworks into practical toolkits, with researchers now exploring its application in medical data analysis, encrypted machine learning, and privacy-preserving data sharing. Despite its computational intensity, HE represents a cutting-edge solution for securing sensitive healthcare data against unauthorized exposure during processing, especially when integrated with blockchain-based storage and access control systems. This unique combination enables decentralized systems to both compute and validate claims without revealing private information at any stage.

*2.4 Related Work*

A number of researchers have explored the integration of blockchain and privacy-enhancing technologies in the context of healthcare data management, though relatively few have specifically targeted the insurance claim domain. Yue et al. [13] presented an early prototype for decentralized medical data management using blockchain, focusing on fine-grained access control and immutable record-keeping. While the system demonstrated blockchain's effectiveness for securing medical records, it did not address the computational aspect of insurance claims, leaving sensitive data exposed during validation. Similarly, Kuo et al. [8] discussed blockchain's potential for enhancing insurance claim workflows but did not combine it with cryptographic computation techniques like HE. This limits the system's ability to maintain end-to-end data privacy during processing.

More recent studies have introduced hybrid solutions combining blockchain with advanced cryptographic techniques such as Secure Multi-Party Computation (SMPC). Sharma et al. [23], for example, proposed a privacy-preserving medical data collaboration system based on blockchain and SMPC, enabling secure joint computations across multiple healthcare entities without direct data sharing. Although SMPC provides strong privacy guarantees, it typically incurs higher communication overhead and requires synchronous cooperation between parties, making it less efficient for real-time insurance claim processing. Similarly, Fan and Xiong [24] explored the use of differential privacy combined with blockchain for healthcare datasets, primarily targeting statistical analysis rather than the detailed computation of individual claims. Another important contribution comes from Chen et al. [25], who combined homomorphic encryption and blockchain for secure genomic data sharing, providing proof of the feasibility of encrypted computation in decentralized medical data scenarios.

Bhandari et al. [26] investigated the convergence of machine learning and blockchain technologies to enhance security and transparency in cyber threat intelligence systems. Their work demonstrated how blockchain's immutable ledger and decentralized validation mechanisms can strengthen data integrity and trust in distributed machine learning applications. Building upon this foundation, Jain et al. [27] revisited fully homomorphic encryption (FHE) schemes for privacy-preserving computation, offering an extensive evaluation of computational efficiency, scalability, and security trade-offs within cloud environments. In a related direction, Adamsetty et al. [28] analyzed the role of homomorphic encryption in securing machine learning classifiers, showing that encrypted inference can maintain high predictive accuracy while preserving data confidentiality—thereby addressing critical concerns in privacy-sensitive AI deployments. Extending cryptographic research into the quantum domain, Savadatti et al. [29] examined quantum fully homomorphic encryption (QFHE) frameworks and proposed a hierarchical memory management architecture designed to improve computational efficiency and scalability in quantum environments. Collectively, these studies illustrate a progressive research trajectory that bridges cryptography, machine learning, and emerging computing paradigms to enhance the security and privacy of intelligent systems.  However, the specific context of medical insurance claims remains underexplored, especially when it comes to end-to-end integration of HE and blockchain to enable both privacy-preserving computation and decentralized auditability. This gap underscores the need for innovative systems like the one proposed in this research, which aims to unify the privacy strengths of homomorphic encryption with the integrity, transparency, and automation features of blockchain to create a secure and efficient medical insurance claim processing framework.

## III. SYSTEM DESIGN & ARCHITECTURE

*3.1 Overview Diagram*

The proposed system is designed to enable secure, transparent, and privacy-preserving medical insurance claim processing by combining the capabilities of homomorphic encryption (HE) and blockchain technology. The architecture comprises six main components: Patients, Hospitals, Insurers, Smart Contracts, the HE Engine, and the Blockchain Network.

- Patients: Individuals seeking medical services who initiate the insurance claim request.
- Hospitals: Authorized healthcare providers responsible for creating, encrypting, and forwarding medical records.
- Insurers: Insurance companies tasked with verifying claims and processing reimbursements based on encrypted data.
- Smart Contracts: Self-executing agreements deployed on the blockchain, programmed to automate the claim processing workflow.
- Homomorphic Encryption (HE) Engine: The core cryptographic module that enables computations to be performed directly on encrypted medical records.
- Blockchain Network: A decentralized ledger (modeled using Ganache for Ethereum simulation) that provides immutable storage, tamper-proof logging, and trustless verification of transaction records.

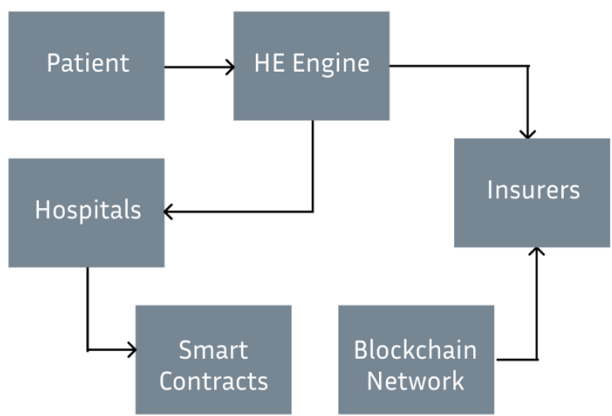

Figure 1 : Key Components of the architecture

Figure 1 Illustrates the distributed architecture that ensures sensitive data remains encrypted during processing and that all interactions are traceable and tamper-resistant, eliminating the need for a trusted central authority.

*3.2 Workflow*

*The system follows a clearly defined, privacy-focused workflow for insurance claim processing:*

1. A patient undergoes medical treatment at a hospital.

2. The hospital encrypts the patient's medical data using the CKKS homomorphic encryption scheme provided by the TenSEAL library.
3. A smart contract is triggered on the blockchain, registering the claim initiation.
4. The insurer receives the encrypted medical data and uses pre-trained encrypted model weights to compute verification logic without decrypting any data.
5. The claim decision, once computed, is recorded immutably on the blockchain, alongside hashed versions of both client data and server responses for integrity verification.
6. When the patient or insurer wishes to access the claim result, the encrypted outcome is decrypted on the client side, and hash-based validation ensures the data has not been tampered with during transit or storage.

This process guarantees that patient data is never exposed in plaintext, neither to the insurer nor any intermediary, while maintaining a verifiable and efficient claim settlement system.

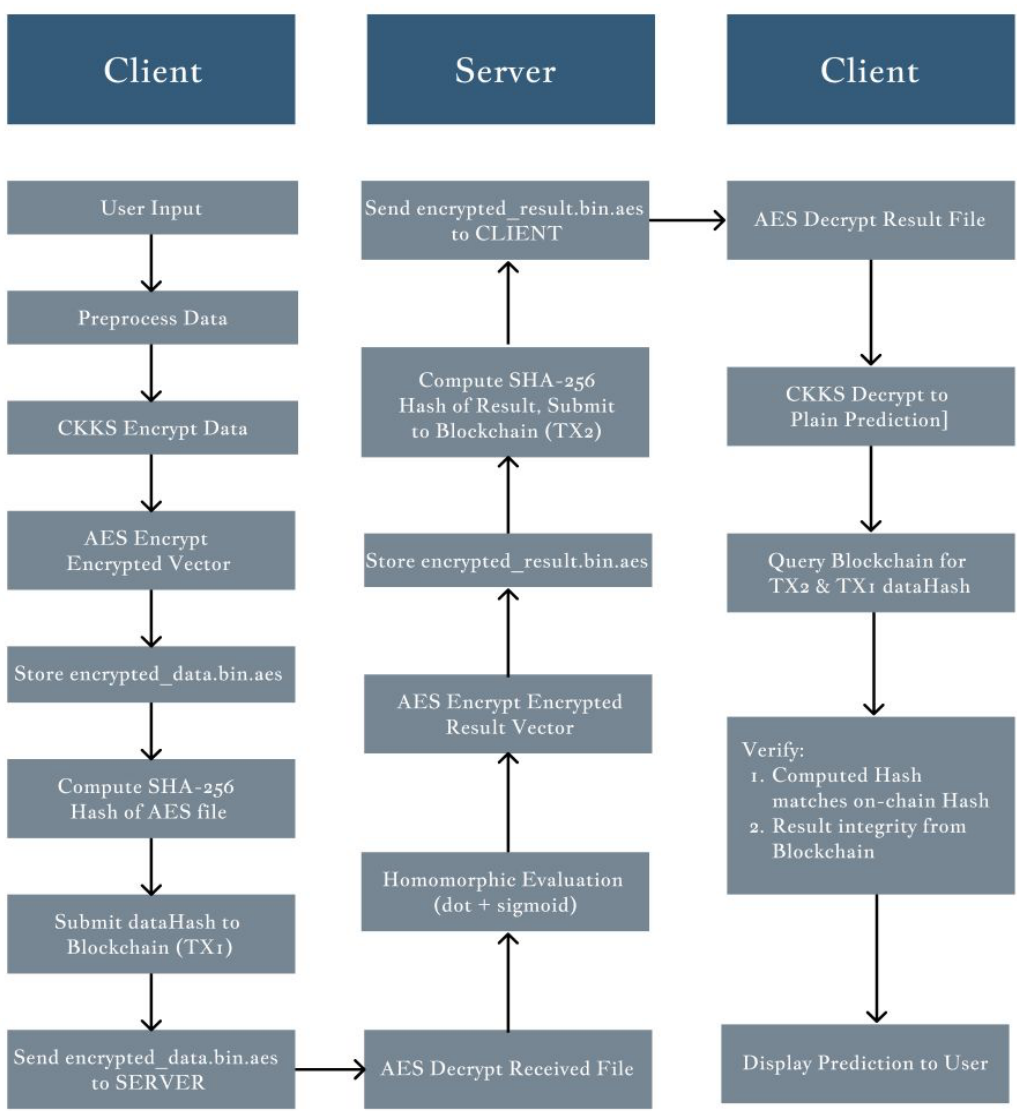

Figure 2 : End-to-end workflow between all the entities

Figure 2 presents a complete end-to-end pipeline for a secure and privacy-preserving machine learning prediction system. It integrates homomorphic encryption (CKKS) for encrypted computation, AES for file-level encryption, and blockchain for ensuring data integrity and verifiability. The entire pipeline is broken into three primary stages—Client, Server, and Client (Post-Prediction).The process involves the client-side encryption of user data using CKKS and AES, secure transmission to the server, homomorphic evaluation on the encrypted data, result re-encryption, and verification via blockchain hashes to ensure data integrity and result authenticity before final decryption and display to the user.

*3.3 Modules*

**Data Encryption Module (HE)**

This module applies homomorphic encryption using the TenSEAL library with the CKKS scheme, which is optimized for operations on approximate real-number data. CKKS was selected over other schemes such as BGV or BFV based on its suitability for floating-point medical datasets and its balance between computation speed and precision. The module ensures that sensitive patient data remains confidential, even while mathematical computations for verification and reimbursement are performed by the insurance provider.

**Smart Contracts for Claim Verification**

Smart contracts are responsible for automating the claim lifecycle on the blockchain. Written using Solidity and deployed on an Ethereum test network via Ganache and Truffle, these contracts manage essential events such as claim submission, verification acknowledgment, and status updates. Their automated nature eliminates human error, reduces administrative overhead, and ensures compliance with pre-agreed logic.

**Blockchain Ledger Management**

This module handles the immutable logging of transaction data including medical claim hashes, encrypted outputs, and decision flags. The distributed and append-only nature of blockchain technology ensures transparency, accountability, and resistance to fraud or tampering, as every transaction can be independently audited.

Figure 3 outlines the complete flow of a data submission to the blockchain. Beginning with user input, the data is hashed using SHA-256 and signed using the user's private key. The signed data is submitted via a smart contract (e.g., addHealthRecord) that logs the data into a new transaction. After broadcasting the transaction, it is appended to a new block in the blockchain ledger. Finally, data retrieval and verification are achieved through audit trails and hash matching. In the context of the work, Figure 3 illustrates the process that ensures traceable, tamper-proof submission of medical claim data to the blockchain. It highlights how each claim is recorded immutably and later verified by comparing hashes during the result validation phase on the client side.

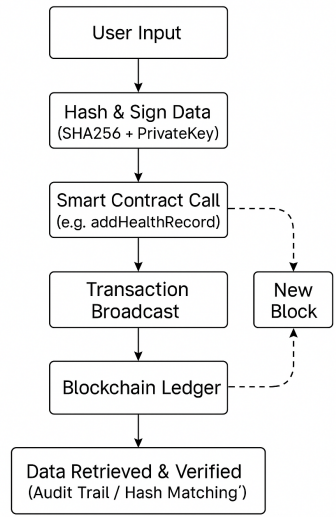

Figure 3: Blockchain Transaction Lifecycle for Medical Data Logging

**Claim Decision Logic**

This component executes the core decision-making process using machine learning models. Initially, the system is trained on unencrypted datasets to compute attribute weights and coefficients. These weights are then encrypted and sent to the server, where claims are verified without decryption. Random Forest, which demonstrated superior performance compared to other models like XGBoost on encrypted data, achieved 85% accuracy under encrypted conditions and 98% accuracy on plain data, highlighting its robustness. The sigmoid function is extremely useful for claim decision logic as it transforms our linear regression output to a probability from 0 to 1. We can then take any probability greater than 0.5 as being 1 and below as being 0. Unlike a stepwise function (which would transform the data into the binary case as well), the sigmoid is differentiable, which is necessary for optimizing the parameters using gradient descent

**Privacy Policy Enforcement**

Beyond technical encryption, the system also enforces privacy policies by embedding them into smart contracts and access control logic. This module ensures that only authorized entities can interact with encrypted datasets or access decryption keys, and that all actions are auditable via the blockchain ledger. It also monitors and logs policy violations or access requests in real-time, providing a systematic way to ensure regulatory compliance and reinforcing patient data sovereignty. In the following, we provide algorithmic representation of the proposed work:

Algorithm: Privacy-Preserving Medical Insurance Claim Processing

Input:
- Patient medical record M$M$
- Insurance policy details P$P$

- Pre-trained ML model weights $W$
- Public/private key pairs for encryption/decryption

Output:
- Secure, auditable claim decision $D$
- Immutable blockchain record of claim and result

Step 1: Data Preparation and Encryption

1.1. Hospital collects and preprocesses patient medical data $M$.

1.2. Hospital encrypts $M$ using Homomorphic Encryption (CKKS scheme):

$$EM \leftarrow HE\_Encrypt(M, public\_key)$$

Step 2: Claim Initiation and Blockchain Logging

2.1. Hospital submits encrypted claim $EM$ to insurer.

2.2. Smart contract is triggered to log claim initiation on the blockchain:

$$Blockchain\_Log(claim\_ID, hash(EM), timestamp)$$

Step 3: Privacy-Preserving Claim Verification

3.1. Insurer receives $EM$ and retrieves encrypted model weights $EW$.

3.2. Insurer computes claim decision on encrypted data:

$$ED \leftarrow HE\_Predict(EM, EW)$$

// All computations are performed without decryption

Step 4: Decision Logging and Integrity Assurance

4.1. Smart contract logs encrypted decision and result hash on blockchain:

$$Blockchain\_Log(claim\_ID, hash(ED), timestamp)$$

Step 5: Result Retrieval and Decryption

5.1. Patient/insurer retrieves $ED$ and associated hashes from blockchain.

5.2. Patient/insurer decrypts $ED$ using private key:

$$D \leftarrow HE\_Decrypt(ED, private\_key)$$

5.3. Patient/insurer verifies integrity by comparing local hash with blockchain hash.

Step 6: Access Control and Auditability

6.1. All access and actions are governed by smart contract logic and access control policies.

6.2. All transactions are auditable and immutable via blockchain records.

The proposed methodology novelty is in the practical, seamless integration of blockchain and homomorphic encryption for secure, privacy-preserving, and fully auditable medical insurance claim processing, with automated smart contract workflows and no exposure of sensitive data at any point in the process. This approach goes beyond existing solutions by ensuring both computational privacy and transparent, tamper-proof record-keeping in a single, unified system.

IV. IMPLEMENTATION

*4.1 Tools & Techniques*

The implementation of the proposed privacy-preserving medical insurance claim processing system relies on a carefully chosen stack of technologies that together ensure data security, transparency, and computational privacy. Table 1 summarizes the core tools and their specific roles within the system architecture.

| Component | Technology | Purpose |
|---|---|---|
| Blockchain Platform | Ethereum (Ganache+ Truffle) | Simulates a decentralized ledger for secure, immutable claim logging and smart contract deployment. |
| Homomorphic Encryption | TenSEAL (CKKS Scheme) | Enables computation on encrypted medical data without decryption. |
| Backend Development | Python | Handles data preprocessing, encryption, server-side logic, and integration workflows. |
| Smart Contract Development | Solidity | Automates claim validation, logging, and decision workflows on the Ethereum blockchain. |

Table 1: Components and Technologies Used in the Insurance Claim Processing System

The selected tools form a secure and efficient pipeline for medical insurance claim processing. Ethereum, along with Ganache and Truffle, ensures tamper-proof storage and automated claim validation through smart contracts. TenSEAL enables encrypted medical data to be processed without decryption using the CKKS scheme, preserving patient privacy. Python handles backend operations like model training, encryption, and system integration, while Solidity automates claim workflows on the blockchain. Together, this stack addresses both data privacy and system transparency challenges.

Table 1 illustrates the core components and technologies employed in the privacy-preserving insurance claim processing system. The blockchain platform utilizes Ethereum, with Ganache and Truffle used to simulate a decentralized ledger and facilitate smart contract deployment for secure and immutable claim logging. Homomorphic encryption is implemented using the CKKS scheme via TenSEAL, enabling computations directly on encrypted medical data without requiring decryption. Python serves as the primary tool for backend development, managing data preprocessing, encryption, server logic, and workflow integration. Finally, Solidity is used for smart contract development to automate claim validation, logging, and decision-making processes on the blockchain.

*4.2 Implementation Steps*

The implementation of the proposed system followed a structured pipeline to integrate blockchain, homomorphic encryption, and machine learning-based claim verification into a unified workflow.

The first step involved data modeling of medical records and claims, where real-world healthcare scenarios were synthetically structured into datasets. Medical records were designed to include multiple attributes such as patient demographics, diagnosis codes, treatment details, and billing amounts. This structured data served as the input for both the machine learning model and the encryption pipeline. The dataset used for this project contains synthetic medical insurance claim records. Each row represents an individual insurance applicant and includes various attributes relevant to claim processing. The key features include demographic details (such as age and sex), lifestyle indicators (such as smoking status and BMI), and regional information. The dataset also contains the insurance **charges** as the target variable, which represents the cost billed for a medical claim. This data serves as the input for preprocessing, encryption, and secure evaluation in the homomorphic encryption and blockchain-integrated claim processing pipeline.

Next, key generation and homomorphic encryption scheme setup were carried out using the TenSEAL library. The CKKS scheme was selected due to its efficiency in handling approximate arithmetic operations, which are especially common in medical and billing data. Public and private key pairs were generated for encrypting and decrypting patient data. The machine learning model was first trained on unencrypted data to compute optimal attribute weights and model parameters. Once the model was trained, its weights were encrypted and sent to the server for inference on encrypted input data.

The third stage focused on blockchain smart contract logic for claims. Smart contracts were designed to automate the claim lifecycle, from registration to result logging. Upon receiving encrypted medical data from the client, the smart contract triggers the verification process, logs the hash of the encrypted data, and later stores the result hash once the processing is completed. These hashes provide an immutable audit trail, ensuring data integrity throughout the process. A crucial part of the system was the integration pipeline between the HE module and smart contracts. The encrypted patient data was transmitted to the server, which used the encrypted model weights for processing without decrypting the data. Once a decision was computed, the result and its hash were logged back onto the blockchain via the smart contract. This enabled patients and insurers to cross-verify the integrity of the processed result against the stored hash, ensuring full transparency and trust without sacrificing privacy.

Finally, the system underwent testing and validation using a combination of both encrypted and unencrypted datasets. The Random Forest model demonstrated high accuracy on both fronts, achieving 98% on unencrypted data and 85% on encrypted data — outperforming models like XGBoost in encrypted conditions. The testing phase also validated the integrity of the entire system, confirming that hash verification and encrypted claim processing were robust against tampering, unauthorized access, and computational errors. The complete implementation flow, including smart contract deployment, encrypted data transmission, server-side processing, result hashing, and client-side result decryption and verification, is documented in both the experimental log and available source code, which can be accessed via the project's GitHub repository.

## V. SECURITY AND PRIVACY ANALYSIS

The proposed system addresses core security and privacy challenges in medical insurance claim processing by integrating homomorphic encryption and blockchain into its design. Each layer of the system plays a critical role in safeguarding sensitive patient information while maintaining transparency and accountability for all stakeholders. Homomorphic Encryption (HE) ensures data confidentiality by allowing computations to be performed on encrypted data without the need for decryption. When medical records are encrypted using the CKKS scheme, insurers are able to verify claims and compute reimbursement logic without ever accessing the underlying plaintext data. This prevents unauthorized disclosure even during processing, ensuring the system remains compliant with privacy regulations such as HIPAA and GDPR, and significantly reducing the risk of data exposure during transmission and storage. Blockchain technology guarantees data integrity and transparency by recording claim-related transactions in an immutable and distributed ledger. Once a claim or its verification result is logged on the blockchain, it cannot be altered without consensus, protecting against tampering or fraud. This distributed structure eliminates single points of failure common in centralized systems and offers all parties — patients, hospitals, and insurers — the ability to independently verify the integrity of the stored data.

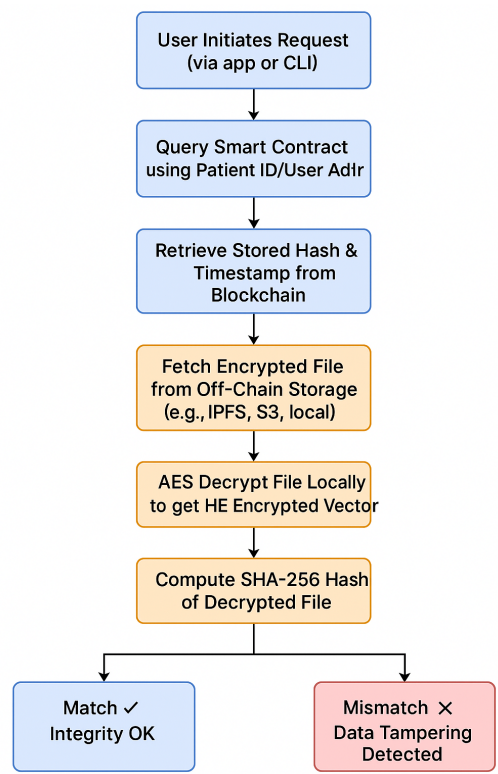

Figure 4: Client-Side Integrity Verification and Tamper Detection Flow

Figure 4 describes the process by which a user (client) verifies the authenticity and integrity of a retrieved encrypted medical file. The client queries the blockchain using the patient ID or user address to obtain the stored SHA-256 hash and timestamp. The corresponding encrypted file is then fetched from off-chain storage (such as IPFS or S3). After AES decryption, the file is converted into its homomorphically encrypted (HE) vector form. A new SHA-256 hash is computed locally and compared to the original stored hash. In the context of the project, figure 4 illustrates the critical step of ensuring end-to-end data integrity during claim result retrieval. It ensures that no tampering has occurred during storage or transmission. A mismatch indicates potential malicious activity or corruption, whereas a match confirms a valid, authenticated medical claim result.

Access Control is enforced through both smart contract rules and encryption. Only authorized entities (such as patients, hospitals, and insurance companies) hold the keys required to encrypt or decrypt data, and smart contracts control the conditions under which data can be logged or processed. This ensures that only verified participants interact with the claim processing pipeline. Auditability is achieved via smart contracts deployed on the blockchain. These contracts automate the verification, status updates, and settlement processes while recording every interaction on-chain. The transparent and verifiable nature of this setup allows for comprehensive audits, giving regulators and stakeholders an immutable history of all claim submissions and decisions. By combining these mechanisms, the system creates a strong security framework that offers end-to-end protection for sensitive medical and insurance data, while maintaining operational transparency and trustworthiness across all participants.

## VI. SECURITY AND PRIVACY ANALYSIS

The system was rigorously tested to assess its performance, privacy guarantees, and usability under real-world medical insurance claim scenarios. Evaluation focused on quantifying encryption overhead, transaction handling on blockchain, and verifying secure claim processing in untrusted environments.

*6.1 Tools & Techniques*

The performance of the system was measured across several dimensions, including data encryption and decryption, encrypted claim processing, smart contract execution, and blockchain storage behavior. The architecture leverages both homomorphic encryption for secure computation and blockchain for data integrity and auditability. These results show that while encryption and blockchain introduce overhead, the system still maintains acceptable efficiency, with secure data handling and fast, immutable logging through smart contracts. Table 2 illustrates the key performance metrics measured during the implementation of the secure insurance claim processing system. Data encryption using the CKKS scheme via TenSEAL takes approximately 0.35 seconds per medical record, while encrypted claim processing on the server side requires around 2.4 seconds per claim. Decryption of the computed results on the client side takes roughly 0.22 seconds. The Ethereum-based smart contract execution for claim validation on the Ganache test network averages about 1.1 seconds. The blockchain demonstrated a transaction throughput of approximately 25 to 30 transactions per second. Additionally, the use of homomorphic encryption and on-chain data logging results in a storage overhead of nearly 2.3 times compared to plaintext data.

| Metric | Measured Value | Description |
|---|---|---|
| Data Encryption Time | ~0.35 seconds per record | Encrypting medical data with CKKS using TenSEAL. |
| Claim Processing Time (Encrypted) | ~2.4 seconds per claim | Server-side computation using encrypted model weights. |

| Data Decryption Time | ~0.22 seconds per result | Decrypting results on the client side. |
| Smart Contract Execution Time | ~1.1 seconds | Transaction validation via Ethereum smart contract on Ganache. |
| Blockchain Transaction Throughput | ~25–30 transactions/sec | Measured on Ganache local Ethereum test network. |
| Storage Overhead | ~2.3x compared to plaintext | Due to homomorphic ciphertexts and hash logging on-chain. |

Table 2: System performance metrics for claim processing.

*6.2 Privacy Evaluation*

The system is built to enforce strict privacy even in untrusted environments. Homomorphic encryption ensures data remains encrypted during computation, and AES safeguards data in transit.

| Test Scenario | Privacy Outcome |
|---|---|
| Encrypted Data Transmission | No leakage (AES encryption) |
| Server-side Computation | No leakage (Homomorphic Encryption) |
| Blockchain Logging | Hashes only, no raw medical data stored. |

Table 3: Privacy assurance across system operations.

This combined use of AES for secure transmission and CKKS for encrypted computation ensures that no sensitive information is exposed at any stage of the claim process. Table 3 presents the privacy outcomes observed across different test scenarios in the system. During encrypted data transmission between client and server, AES encryption ensures that no sensitive information is exposed. Server-side computation on encrypted data is performed using homomorphic encryption, guaranteeing that no data leakage occurs even during processing. For blockchain logging, only cryptographic hashes are stored on the Ethereum ledger, ensuring that raw medical data remains completely hidden while maintaining auditability and integrity.

*6.3 Usability*

A full claim lifecycle was tested to validate usability, including encrypted submission, smart contract-driven verification, result recording, and hash-based integrity verification. The system's end-to-end process follows:

i. Blockchain Server Initialization and Smart Contract Deployment

The project begins with the initialization of the blockchain server and the compilation of the smart contract that governs data integrity and computation validation. This smart contract, written in Solidity, includes essential functions such as logComputation() and verifyComputation(), which record hashes of the encrypted data and result on-chain. After deployment, the contract address is made available to both the client and the server via a shared configuration file. Figure 5 illustrates

18the successful compilation and deployment of the smart contract, confirming readiness for secure blockchain interactions.

Figure 5: Smart Contract Deployment Output via Truffle.

ii. Client-Side Medical Data Encryption and Transmission

Once a user inputs medical data, the client applies preprocessing using a trained pipeline. The data is first encrypted using the CKKS homomorphic encryption scheme to enable privacy-preserving computation. This encrypted vector is then secured further using AES encryption before being stored locally in a file, typically named encrypted_data.bin.aes. The dual-layered encryption ensures that sensitive medical data remains private during both transmission and storage. Figure 6 illustrates the successful transformation of user input into encrypted format using CKKS and AES encryption on the client side.

```
 Blockchain Verification:  Valid
PS C:\Users\91941\SVM> python src/client.py
1. Encrypt data
2. Decrypt result
Enter choice (1/2): 1

=== ENCRYPTION MODE ===

Enter the following details:
Age: 19
Sex (0 = female, 1 = male): 0
BMI: 28
Number of children: 0
Smoker (0 = no, 1 = yes): 0
Region (0 = southwest, 1 = southeast, 2 = northwest, 3 = northeast): 1
Recent medical charges: 1254

 Blockchain TX Hash: 4abcddfdfc6c555f018a11764e4633bab420bb44c8debab3536cd36c2df4600a

Encryption complete. Files saved:
- private_context.bin
- public_context.bin
- encrypted_result.bin.aes
- aes.key
PS C:\Users\91941\SVM> python src/server.py

=== SERVER PROCESSING ===

 Blockchain TX Hash: 18c4fd5ba316c5facd0d8439ab6fc2ce635e18ca699dae57ab3aad9c9ab2b8bb

 Computation complete. Encrypted result saved to src/encrypted_result.bin.aes
PS C:\Users\91941\SVM> python src/client.py
1. Encrypt data
2. Decrypt result
Enter choice (1/2): 2

=== DECRYPTION MODE ===

 Model output (probability): 0.4709
 Claim Denied

 Blockchain Verification:  Valid
```

Figure 6: Client-Side Encryption and Blockchain Transaction Logging.

iii. Hash Logging on the Blockchain Server

Before transmitting the encrypted file to the server for evaluation, the client computes a SHA-256 hash of the encrypted data. This hash is then submitted to the blockchain by invoking the logComputation() function with the second argument set to a null placeholder (e.g., 0x00...00). This serves as a cryptographic fingerprint of the submitted data and prevents unauthorized tampering. Figure 7 illustrates the hash logging process, showing the data hash being successfully stored on-chain in a transaction.

Figure 7: Blockchain Transaction Details from Local Ethereum Testnet.

iv. Secure Server-Side Homomorphic Processing

Upon receiving the AES-encrypted file, the server decrypts it to recover the homomorphically encrypted data vector. Using the model coefficients and intercept—computed during an earlier training phase—the server performs a dot product operation followed by an approximate sigmoid function to produce an encrypted prediction result. This result is then encrypted again using AES before being prepared for return to the client. Figure 8 illustrates the secure homomorphic evaluation process, demonstrating both mathematical operations and AES re-encryption on the server side.

v. Result Hashing and Update on Blockchain

Following the generation of the encrypted prediction, the server computes its SHA-256 hash and calls the logComputation() function again—this time with both the data hash and the newly computed result hash. This immutably binds the prediction to the original data and makes it publicly verifiable. Figure 9 illustrates this step, showing the blockchain updated with both the original data hash and the corresponding result hash within a subsequent transaction.

```
🔗 Blockchain Verification: ✅ Valid
PS C:\Users\91941\SVM> python src/client.py
1. Encrypt data
2. Decrypt result
Enter choice (1/2): 1

=== ENCRYPTION MODE ===

Enter the following details:
Age: 19
Sex (0 = female, 1 = male): 0
BMI: 28
Number of children: 0
Smoker (0 = no, 1 = yes): 0
Region (0 = southwest, 1 = southeast, 2 = northwest, 3 = northeast): 1
Recent medical charges: 1254

🔗 Blockchain TX Hash: 4abcddfdfc6c555f018a11764e4633bab420bb44c8debab3536cd36c2df4600a

Encryption complete. Files saved:
- private_context.bin
- public_context.bin
- encrypted_result.bin.aes
- aes.key
PS C:\Users\91941\SVM> python src/server.py

=== SERVER PROCESSING ===

🔗 Blockchain TX Hash: 18c4fd5ba316c5facd0d8439ab6fc2ce635e18ca699dae57ab3aad9c9ab2b8bb

✅ Computation complete. Encrypted result saved to src/encrypted_result.bin.aes
PS C:\Users\91941\SVM> python src/client.py
1. Encrypt data
2. Decrypt result
Enter choice (1/2): 2

=== DECRYPTION MODE ===

🔍 Model output (probability): 0.4709
❌ Claim Denied

🔗 Blockchain Verification: ✅ Valid
```

Figure 8: Result processing and storing

```
Call Gas Limit
==================
50000000

Chain
==================
Hardfork: shanghai
Id:       1337

RPC Listening on 127.0.0.1:8545
eth_blockNumber
net_version
eth_accounts
eth_getBlockByNumber
eth_accounts
net_version
eth_getBlockByNumber
eth_getBlockByNumber
net_version
eth_getBlockByNumber
eth_estimateGas
net_version
eth_blockNumber
eth_getBlockByNumber
eth_estimateGas
eth_getBlockByNumber
eth_gasPrice
eth_sendTransaction

  Transaction: 0xe6ba2614c9fa4f16511a446e82fa6e045a246b1566d5ca8ff6b247782a7694fd
  Contract created: 0x3ab02029f5f0962e3a90edale133bd6f3a003573
  Gas usage: 221296
  Block number: 1
  Block time: Tue Apr 08 2025 14:32:51 GMT+0530 (India Standard Time)

eth_getTransactionReceipt
eth_getCode
eth_getTransactionByHash
eth_getBlockByNumber
eth_getBalance
```

Figure 9: Hashed result on blockchain server

vi. Client-Side Result Retrieval and Integrity Verification

The client receives the encrypted prediction from the server and decrypts it first with AES, then with CKKS, to obtain the final result. It then re-computes the hash of the encrypted data and compares it with the one retrieved from the blockchain. Additionally, the client invokes the verifyComputation() function to validate the link between the data and result hashes, as well as to ensure the chronological order of blockchain transactions. Figure 10 illustrates this final verification step, where the client performs integrity checks against the blockchain before displaying the prediction outcome.

```
🔗 Blockchain Verification: ✅ Valid
PS C:\Users\91941\SVM> python src/client.py
1. Encrypt data
2. Decrypt result
Enter choice (1/2): 1

=== ENCRYPTION MODE ===

Enter the following details:
Age: 19
Sex (0 = female, 1 = male): 0
BMI: 28
Number of children: 0
Smoker (0 = no, 1 = yes): 0
Region (0 = southwest, 1 = southeast, 2 = northwest, 3 = northeast): 1
Recent medical charges: 1254

🔗 Blockchain TX Hash: 4abcddfdfc6c555f018a11764e4633bab420bb44c8debab3536cd36c2df4600a

Encryption complete. Files saved:
- private_context.bin
- public_context.bin
- encrypted_result.bin.aes
- aes.key
PS C:\Users\91941\SVM> python src/server.py

=== SERVER PROCESSING ===

🔗 Blockchain TX Hash: 18c4fd5ba316c5facd0d8439ab6fc2ce635e18ca699dae57ab3aad9c9ab2b8bb

✅ Computation complete. Encrypted result saved to src/encrypted_result.bin.aes
PS C:\Users\91941\SVM> python src/client.py
1. Encrypt data
2. Decrypt result
Enter choice (1/2): 2

=== DECRYPTION MODE ===

🔍 Model output (probability): 0.4709
❌ Claim Denied

🔗 Blockchain Verification: ✅ Valid
```

Figure 10: Client-Side Decryption and Prediction Result with Blockchain Verification.

The results of the proposed blockchain-integrated, privacy-preserving medical insurance claim processing system demonstrate a strong balance of security, efficiency, and practical usability:

The system achieves high model accuracy, with Random Forest classifiers reaching 98% on plaintext data and maintaining a robust 85% accuracy when operating entirely on encrypted data using homomorphic encryption. Operationally, the framework is efficient, with average encryption times of 0.35 seconds per record, encrypted claim processing completed in about 2.4 seconds per claim, and smart contract execution on the blockchain taking approximately 1.1 seconds. The blockchain infrastructure supports a throughput of 25–30 transactions per second, and the storage overhead introduced by encryption and blockchain logging is moderate at 2.3 times that of plaintext data. From a security and privacy perspective, the system ensures that sensitive patient data is never exposed in plaintext during any stage of processing or storage. All claim data and results are hashed and immutably logged on the blockchain, enabling real-time verification and auditability for all stakeholders. Access to decrypted results is strictly controlled by cryptographic keys, ensuring compliance with privacy regulations. The workflow is fully automated through smart contracts, reducing manual intervention and the risk of fraud or error, while providing a transparent, tamper-proof audit trail for every claim. Compared to traditional centralized systems, this approach offers superior privacy, security, and transparency, with minimal loss in predictive performance and practical, real-world feasibility validated through prototype implementation and testing.

In future, several promising directions can further advance the proposed blockchain-integrated, privacy-preserving medical insurance claim processing system. One important area is the optimization of scalability and computational efficiency, particularly for homomorphic encryption and blockchain operations, to enable real-time processing and support for large-scale, multi-institutional deployments. The framework could be extended to support federated learning and collaborative analytics, allowing multiple insurers or healthcare providers to securely train models on encrypted data without sharing raw patient information. Another key direction is the validation and adaptation of the system for diverse, real-world medical datasets and seamless integration with existing electronic health record (EHR) systems, ensuring interoperability and practical adoption. Enhancing access control through dynamic, attribute-based policies and embedding more flexible privacy rules within smart contracts will allow for context-aware data sharing and compliance with evolving regulations. The system could also be expanded to support cross-chain interoperability and multi-level privacy, enabling secure claims processing across different insurers and healthcare networks with hierarchical access and selective disclosure. Improving user experience through intuitive interfaces and conducting pilot studies in real-world settings will be essential for widespread adoption. Finally, ongoing research will be required to strengthen the system's resilience against emerging security threats, including quantum attacks, and to ensure compliance with future privacy, security, and legal requirements as the healthcare and insurance landscapes continue to evolve.

CONCLUSION

In order to address the persistent issues of privacy, trust, and data integrity in the processing of medical insurance claims, this study offers a decentralized architecture that blends homomorphic encryption with blockchain technology. The suggested solution guarantees end-to-end anonymity while preserving verifiable audit trails and automated claim verification through smart contracts by permitting computation on encrypted data and utilizing the immutability and transparency of blockchain. The implementation findings show that combining CKKS-based homomorphic encryption with Ethereum smart contracts is both feasible and effective, providing strong security without appreciably sacrificing performance. Furthermore, during the claim lifetime, the dual-layered encryption scheme and blockchain logging systems successfully guard against illegal access and data manipulation.

26[27] Jain, Nimish, Aswani Kumar Cherukuri, and Firuz Kamalov. "Revisiting fully homomorphic encryption schemes for privacy-preserving computing." In *Emerging technologies and security in cloud computing*, IGI Global Scientific Publishing, pp. 276-294, 2024.

[28] Adamsetty, Rishitha, Aswani Kumar Cherukuri, and Annapurna Jonnalagadda. "Securing machine learning models: Homomorphic encryption and its impact on classifiers." *Annals of Mathematics and Computer Science,* pp. 141-158, 26, 2025.

[29] Savadatti, Shreya, Aswani Kumar Cherukuri, Annapurna Jonnalagadda, and Athanasios V. Vasilakos. "Analysis of quantum fully homomorphic encryption schemes (QFHE) and hierarchial memory management for QFHE." *Complex & Intelligent Systems* 11, no. 6 (2025): 1-28.
CODE & DATASETS

The software code developed and datasets used for this research are openly available at:

https://github.com/HarshitJain0901/CourseProject-NIS-BITE401L-FacultyGuide-Dr.AswaniKumarCherukur-PrivacyPreserving-Insurance-Claims.git